\documentclass[aps,prl,twocolumn,superscriptaddress,showpacs]{revtex4}
\usepackage{amsfonts,amssymb}
\usepackage{graphicx}

\begin{document}


\title{Kinetics of the Wako--Sait\^{o}--Mu\~{n}oz--Eaton Model of Protein
Folding}


\author{Marco Zamparo}
\email{marco.zamparo@polito.it}
\affiliation{Dipartimento di Fisica and CNISM, Politecnico di Torino,
  c. Duca degli Abruzzi 24, Torino, Italy}
\author{Alessandro Pelizzola}
\email{alessandro.pelizzola@polito.it}
\affiliation{Dipartimento di Fisica and CNISM, Politecnico di Torino,
  c. Duca degli Abruzzi 24, Torino, Italy}
\affiliation{INFN, Sezione di Torino}



\begin{abstract}
We consider a simplified model of protein folding, with binary degrees
of freedom, whose equilibrium thermodynamics is exactly
solvable. Based on this exact solution, the kinetics is studied in the
framework of a local equilibrium approach, for which we prove that (i)
the free energy decreases with time, (ii) the exact equilibrium is
recovered in the infinite time limit, and (iii) the folding rate is an
upper bound of the exact one. The kinetics is compared to the exact
one for a small peptide and to Monte Carlo simulations for a longer
protein, then rates are studied for a real protein and a model structure.
\end{abstract}


\maketitle


Many experimental findings on the folding of small proteins suggest a
strong relationship between the structure of the native state (the 
functional state of the protein) and the
folding kinetics (see e.g.\ \cite{Baker,OnuchicWolynes} and refs.\
therein), and several theoretical models have been proposed which aim
to elucidate the protein folding kinetics on the basis of this
relationship. Among these, three studies appeared in 1999 in the same
PNAS issue \cite{Finkelstein, AlmBaker,ME3}, all of them based on
models with binary (ordered/disordered) degrees of freedom associated
to each aminoacid or peptide bond (the bond between 
consecutive aminoacids). The third of these works \cite{ME3}
is particularly interesting because it is based on a model with
remarkable mathematical properties which make it possible to obtain
exact results. It is a one--dimensional model, with long--range,
many--body interactions, where a binary variable is associated to
each peptide bond. Two aminoacids can interact only if they are
in contact in the native state and all the peptide bonds between them
are in the ordered state. Moreover an entropic cost is associated to
each ordered bond. 

A homogeneous version of the model was first introduced in 1978 by
Wako and Sait\^{o} \cite{WS1,WS2}, who solved exactly the equilibrium
thermodynamics. The full heterogeneous case was later considered by
Mu\~noz, Eaton and coworkers \cite{ME1,ME2,ME3}, who introduced the
single (double, triple) sequence approximations, i.e. they considered
only configurations with at most one (two, three) stretches of
consecutive ordered bonds, for both the
equilibrium and the kinetics. They applied the model to the folding of a
16--aminoacid $\beta$--hairpin \cite{ME1,ME2}, and to a set of 22
proteins \cite{ME3}. The equilibrium problem has been subsequently
studied in \cite{Amos}, with exact solutions for homogeneous
$\beta$--hairpin and $\alpha$--helix structures, mean field
approximation and Monte Carlo simulations. The exact solution for the
equilibrium in the full heterogeneous case was given in
\cite{BP}. Moreover, in \cite{P} it was shown that the equilibrium
probability has an important factorization property, which implies the
exactness of the cluster variation method (CVM)
\cite{Kikuchi,An,TopicalReview}, a variational method for the study of lattice
systems in statistical mechanics. Recently, the model has been used to
study the kinetics of the photoactive yellow protein
\cite{ItohSasai1,ItohSasai2} and the free energy profiles and the
folding rates of a set of 25 two--state proteins \cite{HenryEaton}. It
is also interesting that the WSME model, and the technique developed
in \cite{BP}, have found an application in a problem of strained
epitaxy \cite{TD1,TD2,TD3}.

The ultimate purpose of this class of models being the study of the
kinetics, under the assumption that it is mainly determined by the
structure of the native state, it is worth asking whether the exact
solution developed for the equilibrium can be extended to the
kinetics. Strictly speaking, in the general case an exact solution for the kinetics can not
be achieved. Nevertheless, thanks to the factorization property proved in
\cite{P}, it is possible to devise a local equilibrium approach for
the kinetics which can be proved to yield the exact equilibrium state
in the infinite time limit. It is the aim of the present Letter to
illustrate this approach and its properties, and to discuss its
accuracy and its possible applications.

The WSME model describes a protein of $N+1$ aminoacids as a chain of
$N$ peptide bonds (connecting consecutive aminoacids) that can live in
two states (native and unfolded) and can interact only if they are in
contact in the native structure and if all bonds in the chain between
them are native. To each bond is associated a binary variable $m_{i}$,
$i\in\{1,\ldots,N\}$, with values $0,1$ for unfolded and native state
respectively. The effective free energy of the model (sometimes
improperly referred to as Hamiltonian) reads
\begin{equation}
H_N(m)=\sum_{i=1}^{N-1}\sum_{j=i+1}^{N}
\epsilon_{i,j}\Delta_{i,j}\prod_{k=i}^{j}m_{k}
-RT\sum_{i=1}^{N}q_{i}(1-m_{i})
\label{Hamil} 
\end{equation}
where $R$ is the gas constant and $T$ the absolute temperature.
The first term assigns an energy $\epsilon_{i,j}<0$ to the contact
(defined as in \cite{ME3,BP}) between bonds $i$ and $j$ if this takes
place in the native structure ($\Delta_{i,j}=1$ in this case and
$\Delta_{i,j}=0$ otherwise). The second term represents the entropic
cost $q_{i}>0$ of ordering bond $i$.

A crucial step in the exact solution of the equilibrium problem
\cite{BP,P} is a mapping onto a two--dimensional model through the
introduction of the variables
$x_{i,j}\doteq\prod_{k=i}^{j}m_{k}$ which satisfy the
short--range constraints $x_{i,j}=x_{i,j-1}x_{i+1,j}$ for $1\leq
i<j\leq N$. These can be associated to the nodes of a triangular
shaped portion $\Lambda$ of a two--dimensional square lattice, defined
by $\Lambda=\{(i,j)\in\mathbb{N}^{2}:1 \leq i \leq j \leq N\}$. Let
${\cal{C}}_{\Lambda}$ be the set of all configurations $x$ on
$\Lambda$ that fulfil previous constraints and rewrite the effective
free energy 
(divided by $R T$ and leaving apart an additive constant) in the form
\begin{equation}
H_{\Lambda}(x)=\sum_{(i,j)\in\Lambda} h_{i,j}x_{i,j}.
\end{equation}
The corresponding Boltzmann distribution, which will be denoted by
$p_{\Lambda}^{e}$, has been shown \cite{P} to factor as
\begin{equation}
p^{e}_{\Lambda}(x)=\prod_{\alpha \in \cal{A}}\lbrack
p^{e}_{\alpha}(x_{\alpha})\rbrack ^{a_{\alpha}}.
\label{fact}
\end{equation}
Here $\cal{A}$ is a set of local clusters $\alpha\subset\Lambda$ made
of all square plaquettes ($a_\alpha = 1$), the triangles lying on the
diagonal boundary ($a_\alpha = 1$) and their intersections, that is
internal nearest--neighbour pairs ($a_\alpha = -1$) and single nodes
($a_\alpha = 1$). 
For each cluster $\alpha \in \cal{A}$ we denote by
$x_{\alpha}$ ($x_{\Lambda \setminus \alpha}$) the projection of $x$
onto $\alpha$ ($\Lambda \setminus \alpha$), by $\cal{C}_{\alpha}$ the
set of all configurations on $\alpha$ that are projections of 
configurations on $\Lambda$, and define the cluster 
probability as the marginal distribution
\begin{equation}  
p_{\alpha}^{e}(x_\alpha) \doteq \sum_{x_{\Lambda \setminus \alpha}} 
p_{\Lambda}^{e}(x).
\end{equation}
As a consequence of Eq.\ (\ref{fact}), the equilibrium problem can be
solved exactly \cite{P} by means of the CVM.  Let
${\cal{D}}_{\Lambda}$ be the set of all cluster
probabilities $p$ relative to $\cal{A}$ satisfying the compatibility
conditions $p_{\beta}(x_\beta)=\sum_{x_{\alpha \setminus \beta}}
p_{\alpha}(x_\alpha)$ for $\alpha,\beta\in\cal{A}$ and
$\beta\subset\alpha$.  Since the Boltzmann distribution minimizes the
free energy and factors, restricting the variational principle to
distributions with the same property one finds that
$p^{e}\in{\cal{D}}_{\Lambda}$ is the minimum of the variational
free energy
\begin{equation}
F_{\Lambda}[p] = \sum_{\alpha\in\cal{A}}\sum_{x_\alpha} 
a_{\alpha}[\ln p_{\alpha}(x_\alpha)+h_{\alpha}(x_\alpha)]p_{\alpha}(x_\alpha)
\end{equation}
with respect to $p\in{\cal{D}}_{\Lambda}$. Here $h_{\alpha}$
are defined by $h_{\alpha}(x_{\alpha}) \doteq \sum_{(i,j) \in \alpha} h_{i,j}x_{i,j}$
and it follows that
$H_{\Lambda}(x)=\sum_{\alpha\in\cal{A}}a_{\alpha}h_{\alpha}(x_{\alpha})$. 
This variational approach is not the most efficient way to solve the
equilibrium problem, which is handled in \cite{BP} by the transfer
matrix method. Nevertheless it is a good starting point for a very
accurate treatment of the kinetics. 

Our kinetic problem will be formulated in the framework of a master
equation approach. Denoting by $W_{\Lambda}(x'\to x)$ the transition
probability per unit time from the state $x'$ to $x\not=x'$, we have
to solve
\begin{equation}
\frac{d}{dt}p^{t}_{\Lambda}(x)=\sum_{x'\in{\cal{C}}_{\Lambda}}W_{\Lambda}(x'\to
x)p^{t}_{\Lambda}(x')
\end{equation}
where $W_{\Lambda}(x\to x)$ is such that
$\sum_{x\in{\cal{C}}_{\Lambda}}W_{\Lambda}(x'\to x)=0$.  If the
principle of detailed balance holds, i.e. $W_{\Lambda}(x'\to
x)p^{e}_{\Lambda}(x')=W_{\Lambda}(x\to x')p^{e}_{\Lambda}(x)$, and
$W_{\Lambda}$ is irreducible, then the expected equilibrium is reached.

The above problem is in general not exactly solvable and in order to
overcome this difficulty, we shall assume local equilibrium
\cite{Kawasaki,AdvPhys}, that is we shall assume that, provided the
initial condition $p^{0}_{\Lambda}$ factors according to Eq.\
(\ref{fact}) as the equilibrium probability, the solution
$p^{t}_{\Lambda}$ of the master equation factors in the same way at
any subsequent time. With this simplification the master equation yields 
\cite{Thesis,Prep}, for the cluster probabilities,
\begin{equation}
\frac{d}{dt}p^{t}_{\alpha}(x_\alpha) = \sum_{x'\in{\cal{C}}_{\Lambda}}
W_{\alpha}(x'\to x_\alpha) \prod_{\beta\in\cal{A}}
[p^{t}_{\beta}(x_{\beta}')] ^{a_{\beta}},
\label{app_eq}
\end{equation}
where 
\begin{equation}
W_{\alpha}(x'\to x_\alpha) \doteq \sum_{x_{\Lambda \setminus \alpha}} W_{\Lambda}(x'\to x).
\label{wcluster}
\end{equation}
One can show \cite{Thesis,Prep} that the above evolution preserves the
compatibility conditions between the cluster probabilities and that
the free energy is not increasing in time,
\begin{equation}
\frac{d}{dt}F_{\Lambda}[p^{t}]\leq 0
\end{equation}
(equality holding if and only if $p^{t}=p^{e}$). Since $p^e$ minimizes
the free energy, it follows by Lyapunov's theorem that the exact 
equilibrium probability is recovered in the infinite time limit,
\begin{equation}
\lim_{t\to+\infty}p^{t}=p^{e}.
\end{equation}
It is important to stress that in previous approximations \cite{ME2}
for the kinetics the above condition was not satisfied, and the
behaviour of the free energy was not discussed.

Moreover, denoting by $-k$ ($k>0$) the largest eigenvalue of the
jacobian matrix of the r.h.s. in (\ref{app_eq}) evaluated at
equilibrium, we have that \cite{Prep} for all
$p^{0}\in{\cal{D}}_{\Lambda}$ there exist functions
$R_{\alpha}^{t}$ defined on ${\cal{C}}_{\alpha}$,
$\forall~\alpha\in\cal{A}$, having a finite limit for $t\to +\infty$
and such that
\begin{equation}
p^{t}_{\alpha}=p^{e}_{\alpha}+\mbox{e}^{-kt}R^{t}_{\alpha},
\end{equation}
which allows to define the equilibration rate as $k$. It can also be
shown \cite{Prep} that this approximate equilibration rate is an upper
bound of the exact one, which can be intuitively understood by
observing that the local equilibrium assumption implies that we are
dealing with an evolution in a restricted probability space.

It is also important to observe that, since the cluster probabilities
can be written \cite{P} as linear functions of the expectation values
\begin{equation}
\xi_{i,j}(t) \doteq \langle x_{i,j} \rangle (t) =
\sum_{x\in{\cal{C}}_{\Lambda}} x_{i,j} \prod_{\alpha\in\cal{A}}
[p^{t}_{\alpha}(x_{\alpha})] ^{a_{\alpha}}
\end{equation}
(the probability of the stretch from $i$ to $j$ being native) Eq.\
(\ref{app_eq}) can be rewritten as
\begin{equation}
\frac{d}{dt}\xi(t)=f(\xi(t))\label{eq_m}
\end{equation}
with a suitable function $f$, and the kinetic problem is complete with
an initial condition $\xi(0)$.

A couple of important remarks are in order here. First of all, for the
single ``bond--flip'' kinetics which will be used in the following,
Eq.\ (\ref{eq_m}) turns out to be of polynomial complexity and our
approach yields a reduction of computational complexity which makes
the kinetic problem tractable. This might not be true for other, more
complicated, choices of the transition probability, due to the
summation in Eq.\ (\ref{wcluster}). In addition, we observe that the
whole approach can be reformulated in a discrete time framework
\cite{Prep,Thesis}. 

In order to assess the accuracy of our approach, we have applied it to
the kinetics of the 16 residues C-terminal $\beta$--hairpin of
streptococcal protein G B1 \cite{ME1,ME2,BP}. For such a simple system
it is easy to compute the exact time evolution of $\xi$, denoted by
$\xi^{ex}$, which we compare to our results. Parameters for the
effective free energy are taken from \cite{BP}, $T$ is set to 290 K,
the initial condition is the equilibrium state at infinite temperature
and the transition probability is specified by
\begin{equation}
W_\Lambda(x \to x') = 
\frac{\tau^{-1}}{1+\exp[H_\Lambda(x')-H_\Lambda(x)]}
\label{Glauber}
\end{equation}
if $x$ and $x'$ differ by exactly one bond, that is by the value of a
single $m_i$ variable. Here $\tau$ is a
microscopic time scale , $W_\Lambda(x \to x) =
- \sum_{x' \ne x} W_\Lambda(x \to x')$, and $W_\Lambda(x \to x') = 0$
if $x$ and $x'$ differ by more than one bond.

\begin{figure}
\includegraphics*[height=5cm]{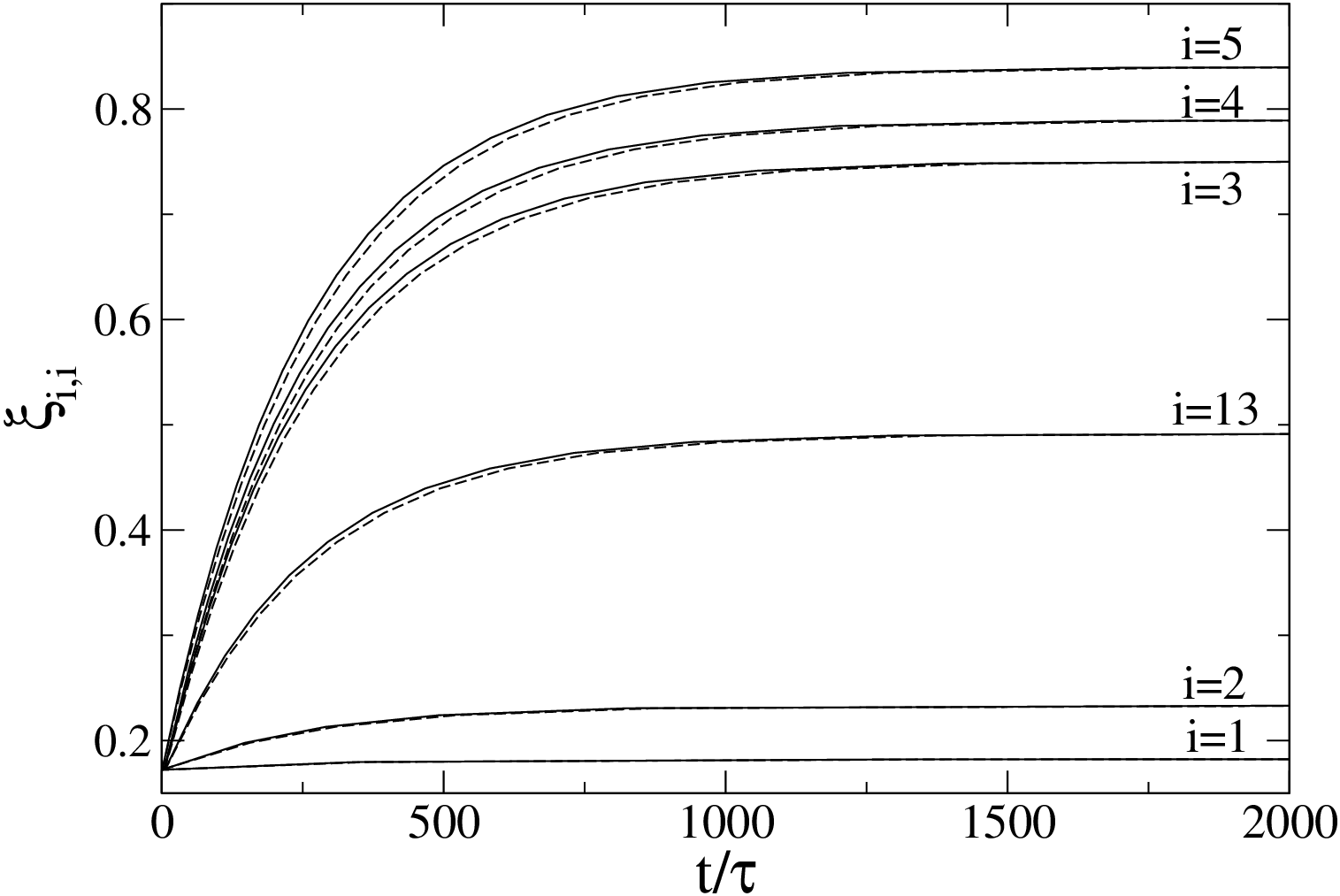}
\caption{\label{fig1} Probability of $\beta$-hairpin bond $i$ being
native, versus time. Solid lines: our results, dashed line: exact results.}
\end{figure}

In Fig.\ \ref{fig1} we report our results for the behaviour of the
probability $\xi_{i,i}$ of bond $i$ being native, as a function of
time. We do not report curves for every value of $i$, since the curves
for $i=5$ to 12 (corresponding to most of the hydrophobic core) are
almost indistinguishable on the graph scale, and in addition
$\xi_{15,15}$ and $\xi_{14,14}$ are almost indistinguishable from
$\xi_{1,1}$ and $\xi_{2,2}$ respectively (which follows from the
symmetry of the hairpin). A more
quantitative measure of the accuracy of our approximation is
\begin{equation}
\max_{t \in \mathbb{R}_{+}} \max_{(i,j)\in \Lambda} 
\vert \xi_{i,j}(t) - \xi_{i,j}^{\rm ex}(t) \vert
\end{equation}
which takes the value $1.6 \cdot 10^{-2}$, which is attained for
$(i,j) = (5,12)$ and $t \simeq 260~\tau$. 
Equilibration rates are also easily computed in our approach. In the
case of Fig.\ \ref{fig1} we obtain $k \simeq 3.93 \cdot 10^{-3}~\tau^{-1}$,
while the exact value is $3.72 \cdot 10^{-3}~\tau^{-1}$.

\begin{figure}
\includegraphics*[height=5cm]{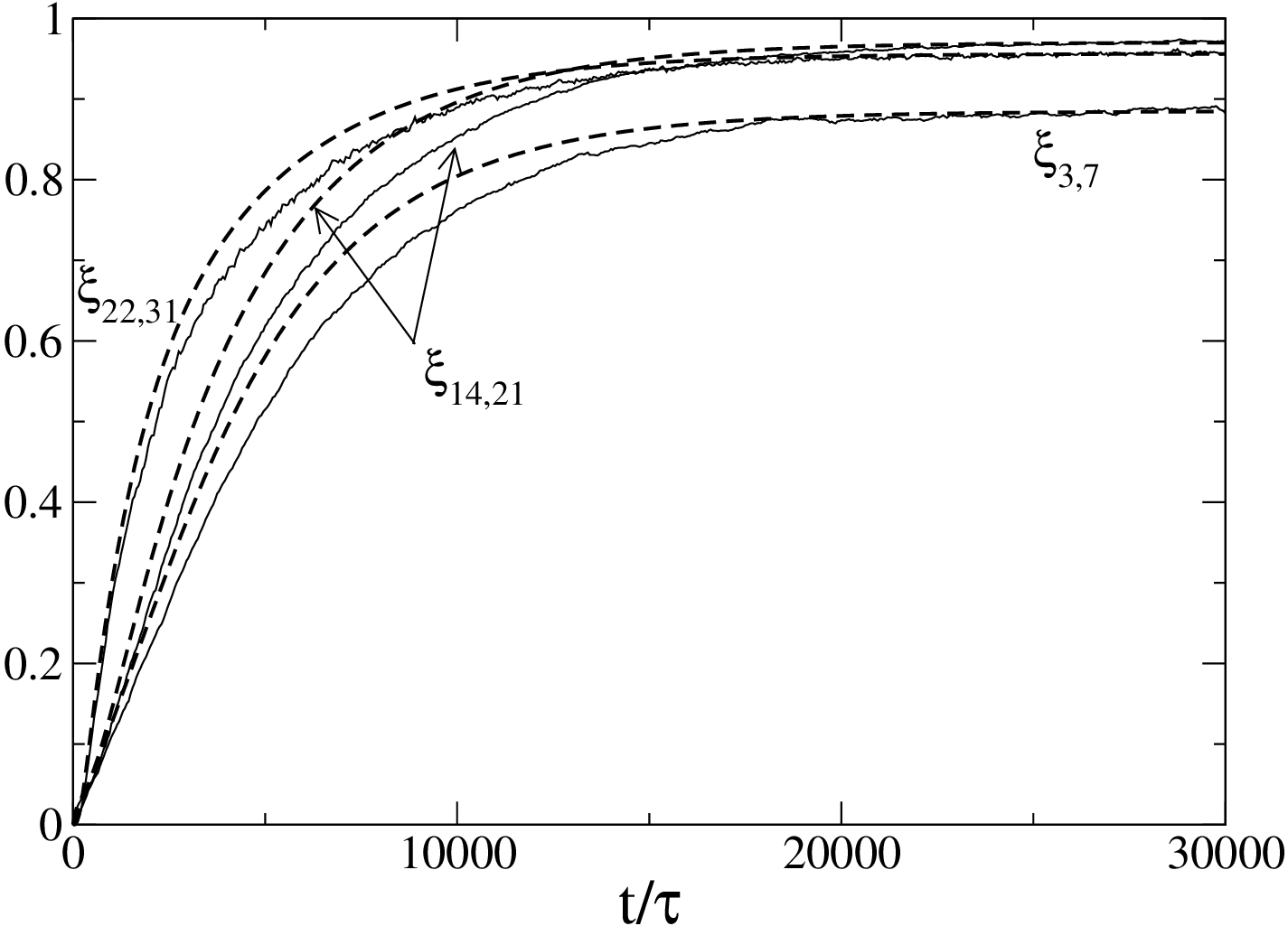}
\caption{\label{fig:1VII} Probability of $\alpha$--helix structures in headpiece subdomain being
native, versus time. Dashed lines: our results, solid lines: Monte Carlo simulations.}
\end{figure}

In order to test our approach on a longer protein, we have considered
the headpiece subdomain of the F-actin-binding protein villin (pdb
code 1VII) and compared our solution with Monte Carlo simulations (in
this case the exact solution is not feasible). The headpiece subdomain
contains three $\alpha$--helices going from bond $3$ to $7$, form bond
$14$ to $17$ and from bond $22$ to $31$. Fig.\ \ref{fig:1VII} shows
the probabilities $\xi_{3,7}$, $\xi_{14,17}$ and $\xi_{22,31}$ versus
time at the temperature of $300$ K.  Parameters for the effective free
energy are taken as in \cite{ME3,BP}, the energy scale is chosen in
order to have $\frac{1}{N} \sum_{i=1}^N \xi_{i,i} = 1/2 $ at
equilibrium at the experimental transition temperature $T = 343$ K
\cite{1VII.bib}. The agreement is still remarkably good and similar
results are obtained for proteins BBL and CI2 (pdb codes 1BBL and 1COA
respectively).
 

\begin{figure}
\includegraphics*[height=5cm]{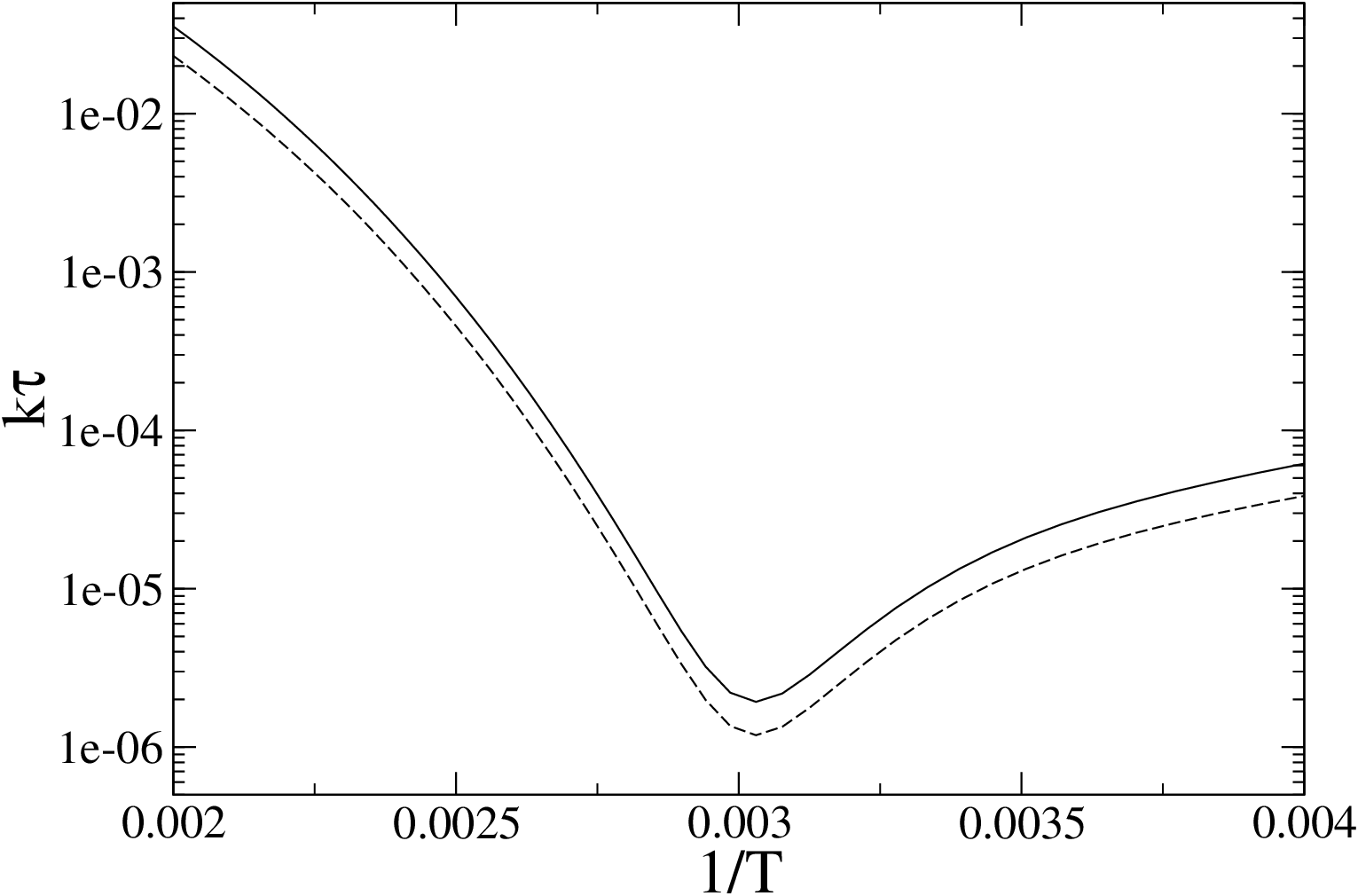}
\caption{\label{fig:PIN1} PIN1 equilibration rate versus inverse
temperature. Solid lines: Glauber kinetics, dashed line: Metropolis
kinetics.}
\end{figure}

Fig.\ \ref{fig:PIN1} reports the equilibration rates as a function of
temperature for the WW domain of protein PIN1 (pdb code 1I6C)
computed using two different choices for the transition probability:
Glauber kinetics, corresponding to Eq.\ (\ref{Glauber}) and  Metropolis
kinetics, where Eq.\ (\ref{Glauber}) is replaced by
\begin{equation}
W_{\Lambda}(x \to x') = \tau^{-1} \exp[H_\Lambda(x) - H_\Lambda(x')]
\end{equation}
if $H_\Lambda(x') \ge H_\Lambda(x)$ and $W_{\Lambda}(x \to x') = \tau^{-1}$
otherwise. 
Here $q_i$ are chosen as in \cite{Cecco}, $\epsilon_{i,j}$ as in \cite{ME3,BP} and
the energy scale in order to have the transition at the experimental temperature 
of $332$ K \cite{Gruebele}. 
It can be seen that the detailed choice of the kinetics affects only 
marginally the behaviour of the equilibration rates, which is 
in very good qualitative agreement with the experimental results
\cite{Gruebele}. The same behaviour is obtained for protein CI2.

\begin{figure}
\includegraphics*[height=5cm]{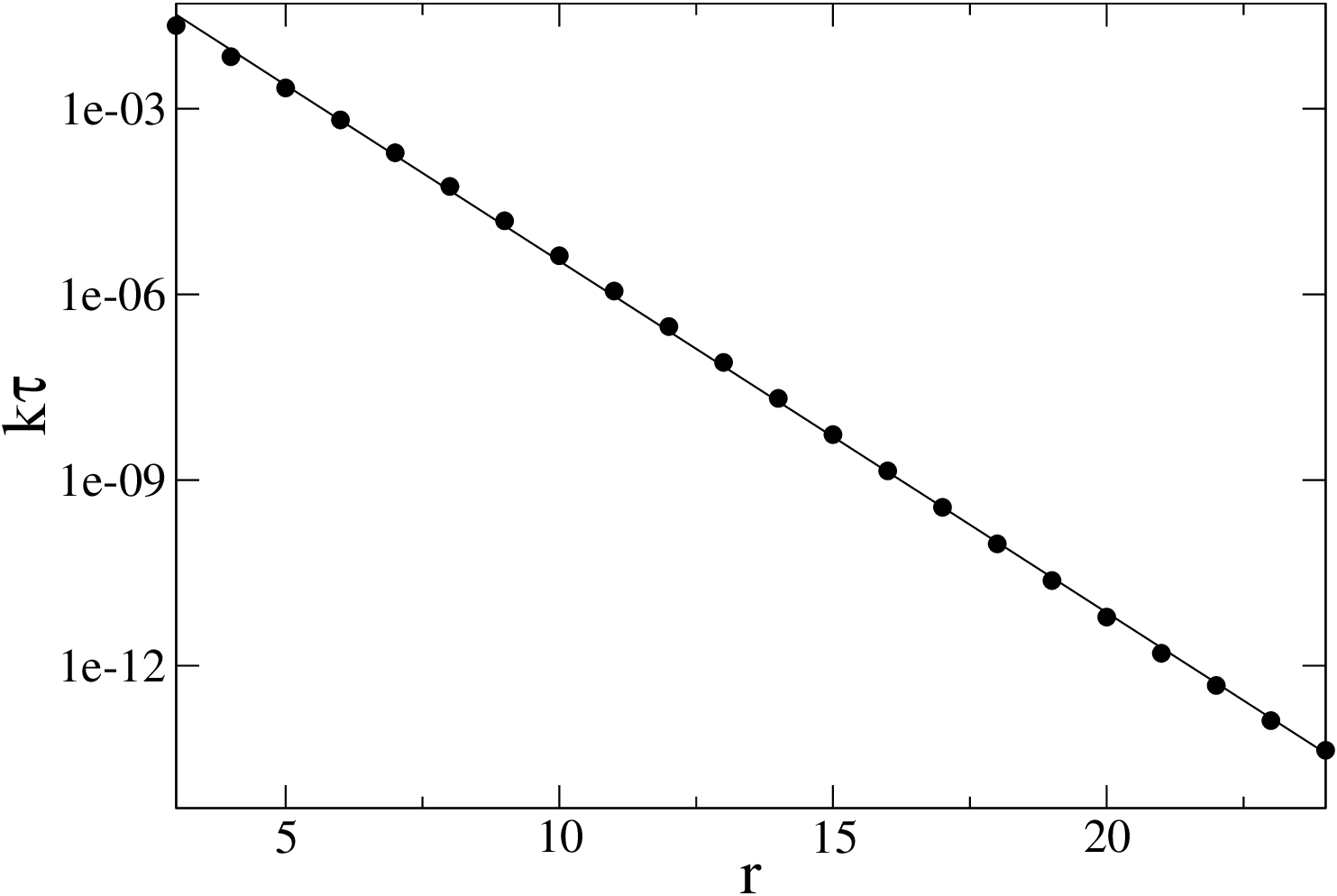}
\caption{\label{fig2} Equilibration rate for a $\beta$--sheet of four
strands versus absolute contact order. Dots: our results, line:
exponential fit.}
\end{figure}

Finally, given the observed correlation between experimental
equilibration rates and structural characteristics like the absolute
contact order ACO $ = \frac{1}{N_c} \sum_{i < j} \Delta_{i,j} (j-i+1)$
\cite{ACO} ($N_c$ is the total number of contacts, $j-i+1$ the
distance in sequence between aminoacids involved in contact $(i,j)$), we have
computed the rates predicted by our approach for a simple model
structure, an antiparallel $\beta$--sheet made of 4 strands, varying
the length $r$ of the strands. In this case the ACO is equal to $r$,
while the relative contact order is a constant. We have used Glauber
kinetics, $q_i = 2 $ and $\epsilon_{i,j}/(RT)$ independent of $(i,j)$
and such that $\frac{1}{N} \sum_{i=1}^N \xi_{i,i} = 1/2$ at
equilibrium.  Results are reported in Fig.\ \ref{fig2}, which shows an
almost perfect linear correlation between $\log k$ and the ACO.

In conclusion, we can say that the local equilibrium approach for the
kinetics of the WSME model gives very accurate results with respect to
the exact ones. It allows to compute equilibration rates, and hence to
explore the relationship between kinetics and structure of the native
state. The detailed evolution is also available, which can be very
useful to study folding pathways. Work is in progress on several
two--state folders and on the effect of mutations. Details of a few
mathematical proofs will be given in \cite{Prep}.

\begin{acknowledgments}
We are grateful for many fruitful discussions with Pierpaolo Bruscolini. 
\end{acknowledgments}


\end{document}